# Effect of NBTI Aging and Process Variations on Write Failures in CMOS and FinFET Flip-Flops

Usman Khalid[1], Antonio Mastrandrea[1], Mauro Olivieri[1]

*Abstract*— The assessment of noise margins and the related probability of failure in digital cells has growingly become essential, as nano-scale CMOS and FinFET technologies are confronting reliability issues caused by aging mechanisms, such as NBTI, and variability in process parameters. The influence of such phenomena is particularly associated to the Write Noise Margins (WNM) in memory elements, since a wrong stored logic value can result in an upset of the system state. In this work, we calculated and compared the effect of process variations and NBTI aging over the years on the actual WNM of various CMOS and FinFET based flip-flop cells. The massive transistor-level Monte Carlo simulations produced both nominal (i.e. mean) values and associated standard deviations of the WNM of the chosen flip-flops. This allowed calculating the consequent write failure probability as a function of an input voltage shift on the flip-flop cells, and assessing a comparison for robustness among different circuit topologies and technologies.

## I. INTRODUCTION

Invasive uninterrupted scaling of CMOS and FinFET technologies to nano-scale level leads to various fallouts such as variability of process parameters and aging due to Negative Bias Temperature Instability (NBTI). While the effects of such issues on *performance* figures like leakage power and delay modeling have been profoundly inquired [1][2][16][18], their influence on the *reliability* of digital designs over the years is being modelled and quantified [4][11][13][12] and there is a general correspondence that reliability be a key topic of future digital integrated circuits [14].

Digital operation reliability is hinged upon noise margins, which have been a prime concept since the foundation of logic circuits [8], as they ascertain the robustness of circuit operation to voltage noise in the circuit nodes. The investigation of noise margin behavior impacted by process variability and aging mechanisms is a prerequisite towards the assessment of circuit-level and system-level design reliability.

In this work, we focus our attention on memory elements (registers) in semi-custom digital design. A transient voltage deviation on a register input during a write operation can result in a wrong stored value, i.e. a permanent upset of the system state, and ultimately in a system operation deviance [8]. Therefore, a remarkable role in digital system reliability is played by the *Write Noise Margins* (*WNM*) of registers. *WNM*s are defined as the maximum tolerance to input dc voltage noise during a write operation to avoid the memory element by capturing a wrong logic value and storing it permanently (write failure). The concept of *WNM* differs from the static noise margin of memory cells, the latter being the maximum internal dc voltage noise tolerance to avoid a memory cell state being flipped during idle mode or read mode, thus measuring the memory cell stability [26].

In this work, we address the occurrence of write failures. The occurrence of timing errors, i.e. the possibility that the stored value – albeit correct – becomes valid too late at a register output, is a different failure condition which is not in the scope of this work.

Our work proposes at evaluating and comparing the effect of process variations and NBTI aging over the years on the *WNM*s and the consequent statistical occurrence of write failures in seven types of flip-flop cells. The analysis proceeds from the statistical characterization of *WNM*s, performed by means of Monte Carlo transistor-level simulations, in order to calculate the probability of write failure as a function of an input voltage shift on the flip-flop cells. The seven analyzed flip-flop cells span from dynamic to pseudo-static and static implementations, based on transmission gates, $C^2MOS$ gates, and TSPC gates; the analysis is carried out for two distinct CMOS and FinFET technologies.

The presentation of the work is organized as follows: Section II reports related research and highlights the novel contributions of the work, Section III presents the exact definitions and methodology, Section IV discusses the resulting collection of data on noise margins statistics, usable as the basis for higher level analyses, and calculates the statistical distribution of the occurrences of write failures for each flip-flop as a function of an input voltage shift.

## II. BACKGROUND AND RELATED WORKS

Process parameter variations has caused several concerns as for digital circuit speed and power consumption. In fact, extensive work has been explored on the impact of variability on propagation delay and leakage power [3][2][17][20][21]. In this work, we address the impact of statistical process variations and NBTI aging on the *WNMs* and consequently on the probability of *write failures* in register cells (flip-flops) used in semi-custom digital design.

---

[1] The authors are with the Dept. of Information Engineering, Electronics and Telecommunications, Sapienza University of Rome, Italy, {khalidu, mastrandrea, olivieri}@diet.uniroma1.it

Our focus is on local cell-level variations in equivalent oxide thickness ($t_{oxe}$), channel width (*W*) and length (*L*) for CMOS cells, and equivalent oxide thickness ($t_{oxe}$)[2], channel length (*L*), fin height ($h_{fin}$) and fin thickness ($t_{fin}$) for FinFET cells. The selected parameter variations have become an integral part of degradation of reliability, specifically while catering with NBTI mechanism [22][23][24].

NBTI is a state-of-the-art degradation factor for the operation of digital circuits. The raising electric field through the oxide thickness generates interface traps in P-type devices and results in unwanted increase in threshold voltage ($V_{th}$) [23] which ultimately affects propagation delay and noise margins. In [5] and [18], investigations have been done on propagation delay for various flip-flops subject to NBTI aging.

A crucial datum obtained by device level research is that NBTI actually depends on process parameters including $t_{oxe}$, *L*, *W* in MOSFETs, and $t_{oxe}$, *L*, $h_{fin}$, $t_{fin}$ in FinFETs, so that process variations actually interfere with the aging statistical impact [7][10][25]. In [9], for CMOS gates, *L, W,* and $t_{oxe}$ are taken into account in estimating the gate aging effects, to conclude that NBTI circuit aging can be severely affected by the magnitude of the process variations. In the FinFET case, [10] shows a variability analysis on $t_{oxe}$, *L*, $h_{fin}$, $t_{fin}$ which result to influence the NBTI reliability degradation effect.

In our work, the following established NBTI equations are used within the simulation tools [22] [23]:

$$\Delta V_{th,IT} \approx e^{\left(-\frac{TITTD}{K.T}\right)} \left[\frac{\varepsilon}{t_d}(V_{GS}-V_{th})\right]^{TITCE} \cdot e^{[TITFD \cdot E_d(V_{GS},V_{DS})] \cdot t^{NIT}} , \quad (1)$$

$$\Delta V_{th,OT} \sim e^{\left[\frac{TOTFD + \frac{TOTTD}{T}}{E_d(V_{GS},V_{DS})}\right] \cdot t^{NOT}} . \quad (2)$$

Eq. 1 and 2 show the physical effect of NBTI in terms of voltage threshold change in interface trapped charges ($\Delta V_{th,IT}$) and oxide trapped charges ($\Delta V_{th,OT}$), involving process parameters along with fitting parameters such as TITTD (Temperature dependent component of interface-trap-inducing threshold voltage degradation), TITCE (Inversion charge exponent for interface-trap-inducing threshold voltage degradation), TITFD (Oxide electric field dependence for interface trap inducing threshold voltage degradation), TOTFD (Oxide electric field dependent component for oxide-trap-inducing threshold voltage degradation) and TOTTD (Temperature dependent component for oxide-trap-inducing threshold voltage degradation). The interested reader should refer to [22][23].

The issue of the reliability of logic memory elements has been always related to the definition of noise margins. In SRAM cells, the criterion of maximum squares in the mirrored VTCs [15] has been applied to quantify nominal *static noise margins* [19], which express the stability of the memory element to internal voltage noise during idle mode, read access and write access. SRAM cell stability has been studied in terms of such static noise margins with respect to process variations [26] and to NBTI aging [6]. In [31], the authors present a voltage control technique for improving BTI-degraded SRAM cell stability. In [32], the authors report on optimizing the energy-delay product in flip-flops affected by NBTI aging. In [4], a work is reported on estimating the probability of *read failure* induced by process variations in SRAM cells. A preliminary subset of results of the proposed work has been presented in [35]. To the best of our knowledge no previous work exists on the characterization and study of write failure probability in flip-flops with respect to variability and aging.

III. METHODOLOGY

*A. Flip-flop cells under analysis*

Fig. 1 – 7 show the schematic circuit design of the flip-flop cells considered in this work. In our analysis, all the circuits are based on a master-slave double latch structure and are established topologies in VLSI design. Other more sophisticated topologies have been proposed in the past, featuring automatic clock-gating, implicit pulsed, explicit pulsed, and differential operating mode [28][29][30][33][34], which are out of the scope of this work. The list of the flip-flops in Fig. 1 – 7, which we also use as reference in the presentation of results, follows:

*A)* Transmission Gate based static FF;
*B)* Transmission Gate based pseudo-static FF;
*C)* Bootstrap Transmission Gate based pseudo-static FF;
*D)* C[2]MOS based pseudo-static FF;
*E)* Bootstrap C[2]MOS based pseudo-static FF;
*F)* C[2]MOS based dynamic FF;
*G)* TSPC N[2]MOS-P[2]MOS based dynamic FF.

The static cells store a bit value permanently, the pseudo-static ones store the value permanently provided that the clock signal is maintained above a certain frequency, and the dynamic implementations need a refresh of the stored value at each clock cycle.

---

[2] Equivalent oxide thickness $t_{oxe}$ is mentioned as $e_{ot}$ in FinFET PTM, however we used $t_{oxe}$ as a common naming convention throughout the paper.

Cells *C* and *E* include "bootstrap" circuits, in which a direct input-output transmission gate path in each latch is used to speed-up the propagation delay of the cell.

### B. Write noise margin definition and characterization

For the purpose of *WNM* assessment, we identify an invalid input voltage, as one that actually results in a write failure, in the flip-flop under analysis. The methodology to characterize the noise margins therefore consists of presenting a voltage value on the flip-flop input pin *D* and then checking the actual data recorded by the cell, by examining the output pin *Q* after the clock edge has occurred. Once we find the input voltage thresholds for correct write operation, $V_{writeH}$ and $V_{writeL}$ for '1' and '0' logic input respectively, the high and low noise margins are represented by the widely known relations:

$WNM_H = V_{DD} - V_{writeH}$;   $WNM_L = V_{writeL}$

where we have recalled that output nominal voltage values in CMOS and FinFET cell are $V_{OH} = V_{DD}$ and $V_{OL} = 0V$.

Fig. 8 reproduces SPICE simulation results (for 16 nm CMOS, Flip-Flop A) that exemplify the procedure to measure $V_{writeL}$ and $V_{writeH}$. For different values of actual input voltage V(D) on the clock edge, the output voltage either switches to the correct value or remains practically unchanged.

In the actual $V_{writeL}$ ($V_{writeH}$) characterization procedure, voltage values on node *D* ranging from 0 V to $V_{DD}$ are iterated by a bisection search, until the highest (lowest) voltage on *D* producing a low (high) logic output *Q* is found. The stop criterion of the bisection search algorithms for determining the target $V_{writeL}$ and $V_{writeH}$ values is set to an accuracy of $0.001 \cdot V_{DD}$. The following pseudo code shows the procedure of estimation of $V_{writeL}$ and one can compute $V_{writeH}$ in the same manner.

```
Pseudo code: VwriteL estimation
1: Input: Transistor-level Flip Flop Netlist with Y, Tox, L, W, Vd parameters
2: For each NBTI_Aging_year Y in [0, 1, 2,…, 10]
3:    Update_MOSRA_Aging_value(Y);
4:    For each I in [0…10E4]
5:         Random_value_selection(Tox, L, W);
6:         do
7:              Update input Vd according to bisection search;
8:              Create_Netlist(Tox, L, W, Vd);
9:              Simulate SPICE Netlist;
10:             Extract output VQ value;
11:        While (VQ is not correct);
12:        Output: Save VwriteL = Vd value for Tox, L, W, value;
13:   end For
14: end For
```

### C. Impact of slack time and clock frequency

The setup time of input *D* with respect to the clock edge assumed in the characterization procedure has an impact of the resulting $V_{writeL}$ and $V_{writeH}$. In general, a large setup slack favors wider noise margins, while a slack time close to zero leads to narrower actual noise margins. Therefore, our analysis has been divided into two operating cases:

- long setup slack time, reflecting design cases where the registers (flip-flops) are not on the critical path of the architecture, or the architecture is clocked at low speed, resulting in a large slack. The long setup slack time was fixed at 2 ns for all flip-flops types. We verified that longer slack times do not produce any relevant change.
- zero setup slack time, reflecting the case of registers on the critical path of a high-clock-speed architecture design. Zero slack time models a timing critical situation at the *input* of the flip-flop, possibly producing a logic error in the stored data; it does not refer to a timing error on the output of the flip-flop. In the noise margin analysis at zero setup slack time, the interval between the rising/falling edge of *D* and the rising clock edge was set equal to the minimum setup time for each of the flip-flops.

A potentially important point was the influence of the clock period duration. In fact, in the simulation script to identify $V_{writeL}$ and $V_{writeH}$, we considered the output *Q* correct if the voltage value at node *Q* stably reached the correct nominal value (0 V or $V_{DD}$) within the current clock cycle. Our simulations always showed that, in this respect, the outcome and the consequently extracted noise margins are *not* impacted by the clock period duration for clock frequencies ranging from 10 MHz up to 1 GHZ and above, i.e. in a range typically used in semi-custom designs in the target technologies. All the presented results have been obtained at the reference clock frequency of 1 GHz.

### D. Aging effect and process variation effect evaluation

For the case of CMOS technology, we can formally express the $V_{writeL}$ and $V_{writeH}$ values of a generic flip-flop instance as a pair of functions:

$$V_{writeL} = V_{writeL}(Y, t_{oxe}, L, W), \quad V_{writeH} = V_{writeH}(Y, t_{oxe}, L, W).$$

where $T_{oxe}, L, W$ are to be considered random variables, characterized by technology-dependent nominal value (mean) and standard deviation, and $Y$ is the circuit life in years.

For the case of a generic FinFET-based flip-flop instance, we can similarly write:

$$V_{writeL} = V_{writeL}(Y, t_{oxe}, L, h_{fin}, t_{fin}), \quad V_{writeH} = V_{writeH}(Y, t_{oxe}, L, h_{fin}, t_{fin}).$$

where $t_{oxe}, L, h_{fin}, t_{fin}$ are random variables and $Y$ is the year count.

We implemented the characterization in two phases. First, we targeted the impact of aging $Y$ on the noise margins in nominal process conditions by NBTI simulation analysis. Secondly, we targeted the analysis of NBTI aging together with process variation effects, implemented by Monte Carlo simulations with random variations in the technology model parameters, at each aging step $Y$.

*E. Characterization setup*

All the circuits include an inverter load cell sized 4 times the minimum size (FO4 load). The target technology models are a bulk FinFET 16 nm LSTP process, at nominal 0.85V supply $V_{DD}$, and a bulk CMOS 16 nm HP process, the latter with scaled-up supply $V_{DD}$ from nominal 0.7V to 0.85V in order to compare the resulting noise margins at same supply voltage.

The NBTI aging analysis was performed on the built-in HSPICE tool MOSFET Model Reliability Analysis (MOSRA). According to the typical usage, the simulation was divided into pre-stress analysis (fresh simulations or at zero time) and post-stress analysis (in long time periods) [23]. We chose the post-stress analysis for 10 years with 2 years' time period.

For the zero setup slack time case, we had to *a priori* identify the minimum setup time of each flip-flop type. The state of the art practical definition of minimum setup time has been extensively reported in literature as the setup time causing an increase of 10% of Clock-to-Q nominal propagation delay[3] [9]. Table I reports the characterization obtained for the timing performance of each flip-flop cell in nominal process conditions, obtained by SPICE simulations with MOSRA aging models [23] for different aging times. Note that the minimum setup time degrades (i.e. increases) with aging, so that to provide for 10 years' circuit operating life we assumed the setup time slack at 10 years aging as design reference.

For statistical variability analysis, according to a typical setup in the literature [17], we implemented HSPICE Monte Carlo simulations with normal distributions of the selected process parameters, assuming $3\sigma/\mu = 10\%$. Other statistical distributions obtained from industrial process development kits (PDK) can be applied without any modification to the methodology. The nominal (mean) values and standard deviation values are reported in Table II. Note that for CMOS $L$ and $W$, the nominal values are of *Lint* ($\Delta L$) and *Wint* ($\Delta W$) respectively; however, the standard deviation value 0.26 nm is according to $L$ and $W$ actual value i.e. 16 nm. The relationship between $L$, $W$ with *Lint*, *Wint* is according to BSIM equation i.e. $L_{eff} = L_{drawn} + XL - 2\,L_{int}$ and $W_{eff} = (W_{drawn}/NF) + XW - 2\,W_{int}$ respectively [27]. The reason to consider this assumption is due to unavailable direct values of $L$ and $W$ in CMOS PTM which were needed for simplest SPICE Monte Carlo implementation. However, in FinFET SPICE Monte Carlo implementation there was no assumption needed, as one can directly modify $h_{fin}$, $t_{fin}$ and $L$ in FinFET PTM. For each cases of age $Y$, Monte Carlo simulation has been run for over $10^4$ iterations. The validity of the statistical results has been verified by matching the resulting mean values of noise margins with their corresponding nominal values.

IV. RESULTS

*A. Noise margin distributions*

The statistical distributions of all the noise margins samples obtained by MC simulation showed a good matching with normal distributions. In all cases the substantial matching between nominal values and statistical mean values confirm the correctness of the Monte Carlo simulations. Fig. 9-15 present the results on the nominal (i.e. mean) value and of standard deviation for $V_{writeL}$ and $V_{writeH}$, referring to stress time $Y = 0$ up to $Y = 10$ years. An interesting outcome is that the nominal value of the noise margins is differently affected by NBTI aging in large slack conditions with respect to zero-slack conditions. In large slack conditions the noise margin aging effect behaves asymmetrically, showing nominal $V_{writeH}$ lowering (i.e. improving) and nominal $V_{writeL}$ lowering as well (i.e. degrading). In zero-slack conditions, the aging effect impact is reduced, but both noise margins tend to degrade (i.e. $V_{writeH}$ raising, $V_{writeL}$ lowering). As for the standard deviation values, the results show different NBTI aging impact depending on the Flip-Flop cell, but a general result is that for the FinFET cells the standard deviation values of the noise margins are substantially lower than in their CMOS counterparts, with a practically negligible NBTI aging impact over 10 years. The latter result makes it interesting to investigate the performance of FinFET vs CMOS cells in terms of estimated write failure probability.

*B. Calculation of the probability of write failures*

The calculation of the cumulative normal distribution, based on the obtained values of mean and standard deviation of the *WNM*s, allows us to obtain a picture of the probability of an actual write failure caused by a voltage shift $\Delta v$ at the register input.

---

[3] Technical documents report 5% as well as 10%, depending on the target degree of tolerance assumed for the cell library.

The voltage shift $\Delta v$ can also be seen as the minimum needed noise margin in given application [4]. In fact, the probability that a register stores a wrong bit is the probability that the voltage noise margin is smaller than the input voltage shift $\Delta v$, i.e. formally:

- Logic value '0' at register input:
$$\Pr\{\text{failure}\} = \Pr\{WNM_L < \Delta v\} = \Pr\{V_{writeL} < \Delta v\} = \int_{-\infty}^{\Delta v} \mathcal{N}(x, \mu_{VwriteL}, \sigma_{VwriteL}) \, dx \; ;$$

- Logic value '1' at register input:
$$\Pr\{\text{failure}\} = \Pr\{WNM_H < \Delta v\} = \Pr\{V_{writeH} > V_{DD} - \Delta v\} = \int_{V_{DD}-\Delta v}^{+\infty} \mathcal{N}(x, \mu_{VwriteH}, \sigma_{VwriteH}) \, dx \; .$$

Here, $\Delta v$ is a positive quantity and $\mathcal{N}(x, \mu, \sigma)$ is the normal probability density function with mean $\mu$ and standard deviation $\sigma$.

Fig. 16 shows the results of the above calculation for the zero slack time case, allowing a direct comparison of the flip-flop cells with respect to their robustness to a certain amount of input voltage drop/rise $\Delta v$ during a write operation, as impacted by process variations and NBTI aging.

From our results, while in nominal conditions the CMOS cells outperform FinFET ones as for robustness to input voltage shift (even at 10 years' aging), when considering process variations the outcome is reversed: the larger standard deviations of noise margins in CMOS cells lead to a smoother behavior of the failure probability curves (Fig. 16) and ultimately reduce the yield associated to a certain input noise tolerance more than it happens in FinFET cells. It must be considered, however, that this result only refers to the case $3\sigma/\mu = 10\%$ assumed in both technologies for the selected geometric process parameters.

In general, we observe an asymmetrical performance for low and high logic input values, depending on the cell. This effect is evidenced for both technologies, and suggests that a proper transistor sizing could balance robustness between '1' and '0' logic write operations.

A final most important outcome of the reported analysis, is the identification of the best performing flip-flop circuit topologies with respect to write failures subject to process variations and NBTI aging. In this respect, the validity of the results is enforced by the fact that in both technologies we have the same outcomes. First, bootstrap transmission gate results to enhance robustness to process variations and also to aging, as shown by Fig. 16 for flip-flops (C) and (E). Secondly, flip-flops (G), (F) and (D) appears to be more impacted by NBTI aging than others. Somehow surprisingly, the fully static flip-flop (A) shows a poor performance especially referring to the 10-year aging condition, not very differently from fully dynamic implementations (F) and (G) which in fact were expected to be weak. Overall, the best performance is attained by flip-flop (E), which may be considered for reliability-oriented libraries of semi-custom cells.

## V. CONCLUSIONS

The characterization of *WNM*s of different flip-flop cells have been quantified with respect to the combined effect of NBTI aging and process variations, in a CMOS and a FinFET technology, according to state of the art technology models; the distribution of write failure probability has been also calculated accordingly.

The results show that the NBTI aging affects the noise margins differently in large slack time and zero-slack time conditions. Also, for large slack time an evident asymmetry exists between the aging effects on the logic high noise margin and the logic low noise margin.

The evidently smaller standard deviation in FinFET cells with respect to CMOS, results in better performance as for write failure probability at a given input voltage noise.

As for the cell topology comparison, fully static cells results to perform poorly with respect to pseudo-static ones, and bootstrap transmission gates result to improve the noise margins. Overall, the best performing flip-flop for reliability results to be the bootstrapped C$^2$MOS pseudo-static cell.

ACKNOWLEDGEMENT

The authors are thankful to Sapienza University of Rome for providing the financial resources for this research work.

Table I – Characterized propagation delay and minimum setup time for different types of flip flops at different aging times

| | | MOSFET 16 nm | | | | FinFET 16 nm | | | |
|---|---|---|---|---|---|---|---|---|---|
| | | 0 years | | 10 years | | 0 years | | 10 years | |
| | Q edge | $t_{CKtoQ}$ (ps) | $t_{Setup\ MIN}$ (ps) | $t_{CKtoQ}$ (ps) | $t_{Setup\ MIN}$ (ps) | $t_{CKtoQ}$ (ps) | $t_{Setup\ MIN}$ (ps) | $t_{CKtoQ}$ (ps) | $t_{Setup\ MIN}$ (ps) |
| flip flop A | 0-to-1 | 05.70 | 10.00 | 06.59 | 12.62 | 06.78 | 06.83 | 08.05 | 07.68 |
| | 1-to-0 | 09.38 | 07.04 | 10.34 | 07.66 | 08.38 | 05.58 | 08.44 | 06.17 |
| flip flop B | 0-to-1 | 05.48 | 13.16 | 06.44 | 15.98 | 06.09 | 06.19 | 06.17 | 08.08 |
| | 1-to-0 | 07.96 | 07.27 | 08.35 | 08.29 | 07.34 | 05.16 | 07.37 | 10.01 |
| flip flop C | 0-to-1 | 11.09 | 10.59 | 15.05 | 10.68 | 06.03 | 06.36 | 08.03 | 06.84 |
| | 1-to-0 | 07.30 | 09.20 | 08.80 | 10.35 | 06.08 | 06.13 | 07.46 | 07.01 |
| flip flop D | 0-to-1 | 08.63 | 17.41 | 09.84 | 21.37 | 08.46 | 11.20 | 09.41 | 12.11 |
| | 1-to-0 | 08.58 | 19.79 | 10.04 | 22.17 | 08.70 | 10.22 | 08.80 | 11.58 |
| flip flop E | 0-to-1 | 06.73 | 21.17 | 08.41 | 24.37 | 04.26 | 14.50 | 05.74 | 15.56 |
| | 1-to-0 | 05.98 | 21.77 | 06.66 | 29.00 | 04.68 | 13.47 | 05.55 | 15.45 |
| flip flop F | 0-to-1 | 01.59 | 09.00 | 02.79 | 12.00 | 00.82 | 08.00 | 02.32 | 11.00 |
| | 1-to-0 | 01.95 | 08.00 | 02.70 | 11.00 | 02.06 | 05.00 | 02.83 | 08.00 |
| flip flop G | 0-to-1 | 04.97 | 03.50 | 05.74 | 07.00 | 03.07 | 03.00 | 04.23 | 05.00 |
| | 1-to-0 | 08.94 | 09.60 | 14.40 | 11.98 | 07.15 | 06.00 | 08.26 | 08.00 |

Table II – Process parameters with their mean and standard deviation values

| Parameters | MOSFET 16 nm | | | | | | FinFET 16 nm | | | | | | | |
|---|---|---|---|---|---|---|---|---|---|---|---|---|---|---|
| | $t_{OXE}$ [nm] | | $L$ [nm] | | $W$ [nm] | | $t_{OXE}$ [nm] | | $h_{FIN}$ [nm] | | $t_{FIN}$ [nm] | | $L$ [nm] | |
| | μ | σ | μ | σ | μ | σ | μ | σ | μ | σ | μ | σ | μ | σ |
| N-Type | 0.95 | 0.0317 | 1.45 | 0.26 | 5 | 0.26 | 1 | 0.0334 | 26 | 0.867 | 12 | 0.40 | 20 | 0.667 |
| P-Type | 1 | 0.0334 | 1.45 | 0.26 | 5 | 0.26 | 1 | 0.0334 | 26 | 0.867 | 12 | 0.40 | 20 | 0.667 |

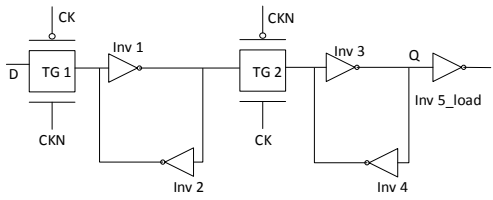

Fig. 1. Transmission Gate based static flip-flop A

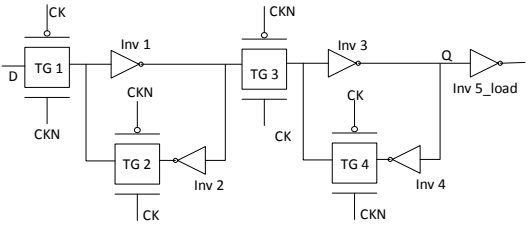

Fig. 2. Transmission Gate based pseudo-static flip-flop B

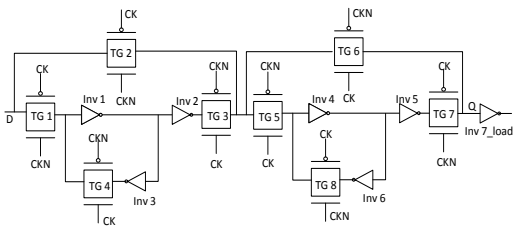

Fig. 3. Bootstrap transmission gate based pseudo-static flip-flop C

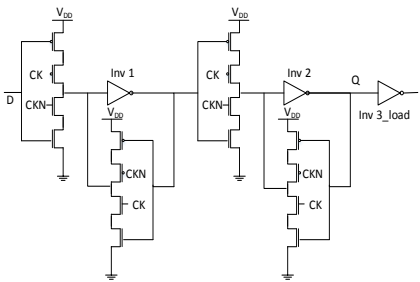

Fig. 4. C$^2$MOS based pseudo-static flip-flop D

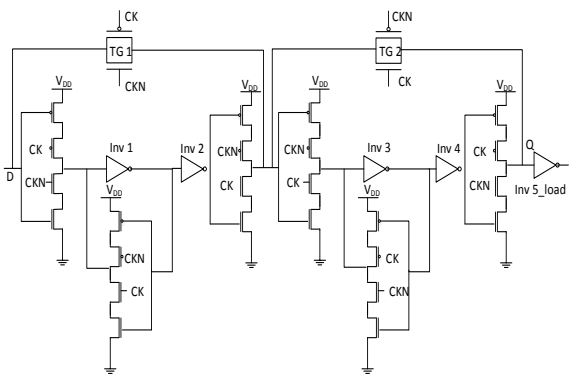

Fig. 5. Bootstrap C$^2$MOS based pseudo-static flip-flop E

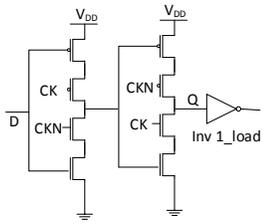

Fig. 6. C²MOS based dynamic flip-flop F

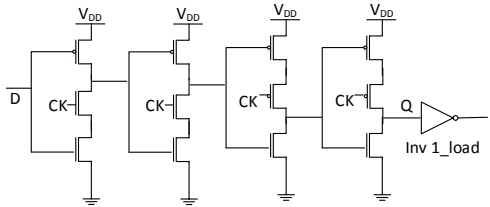

Fig. 7. TSPC N²MOS-P²MOS based dynamic flip-flop G

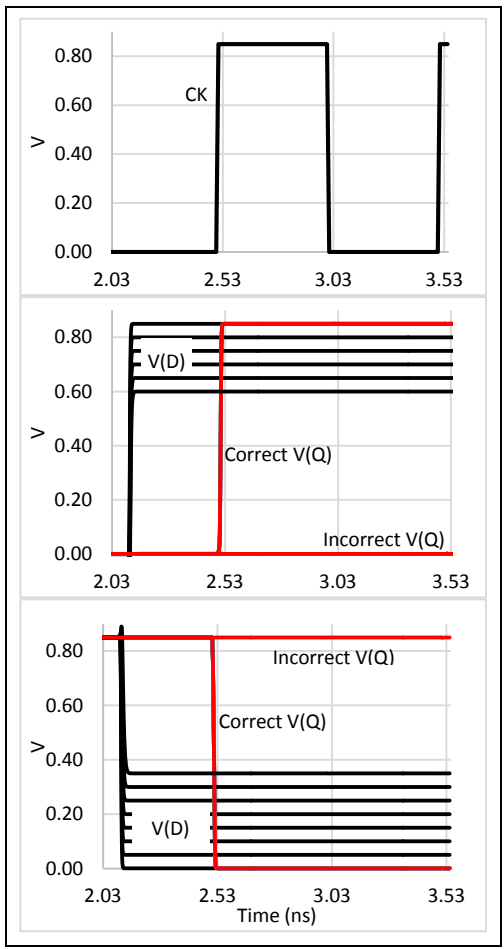

Fig. 8. Sample of SPICE simulations (16 nm CMOS, FF-A). Top: Clock, Mid: $V_{writeH}$ characterization, Bottom: $V_{writeL}$ characterization

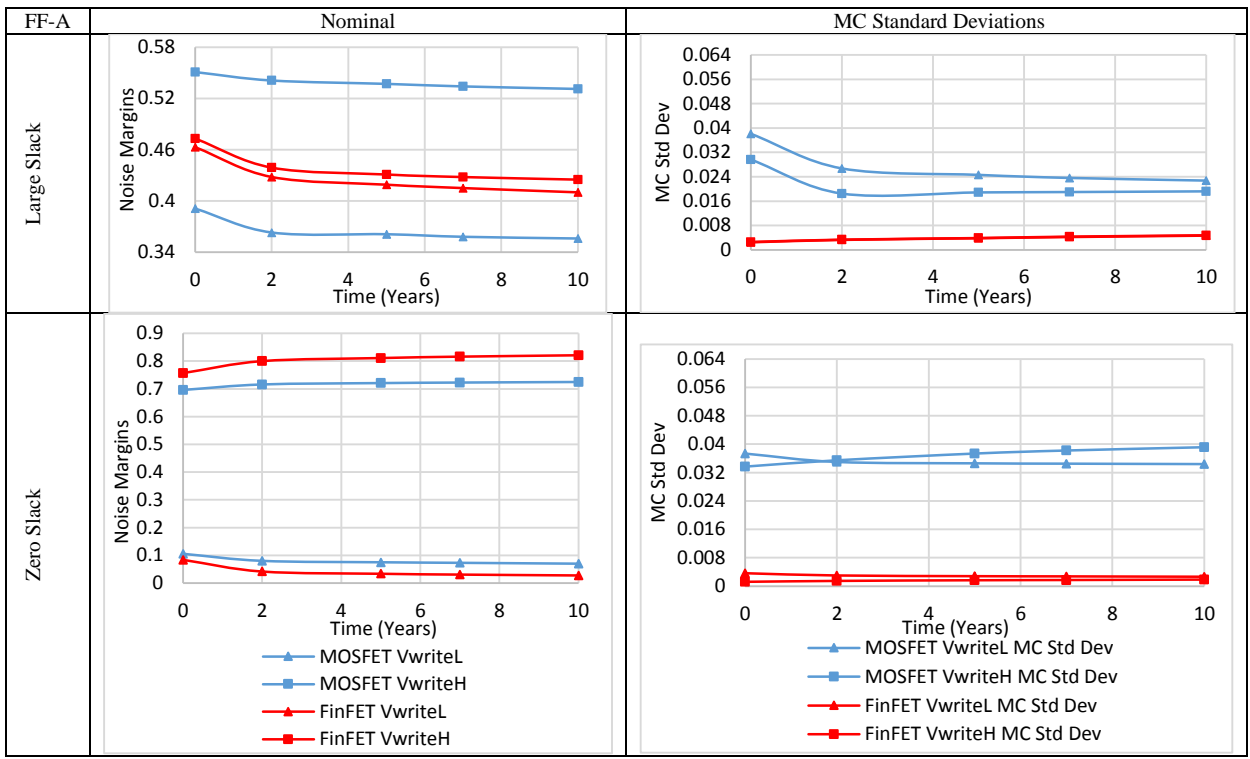

Fig. 9 $V_{writeH}$, $V_{writeL}$ nominal values and standard deviations for flip-flop-A

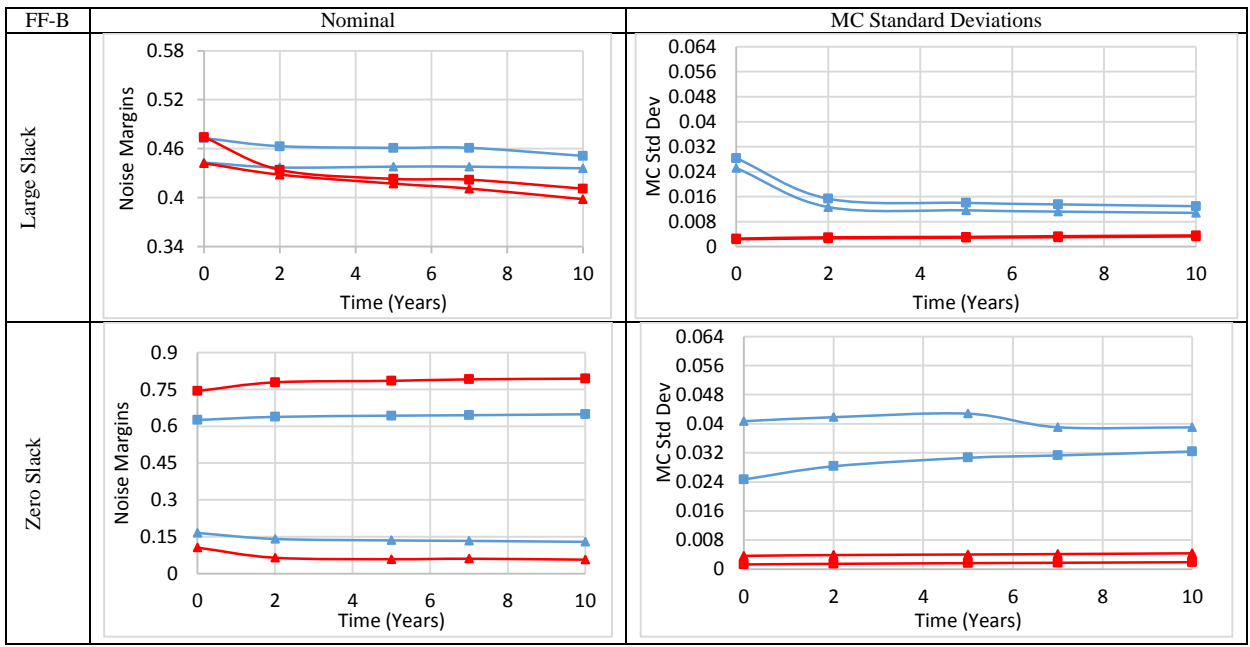

Fig. 10 $V_{writeH}$, $V_{writeL}$ nominal values and standard deviations for flip-flop-B

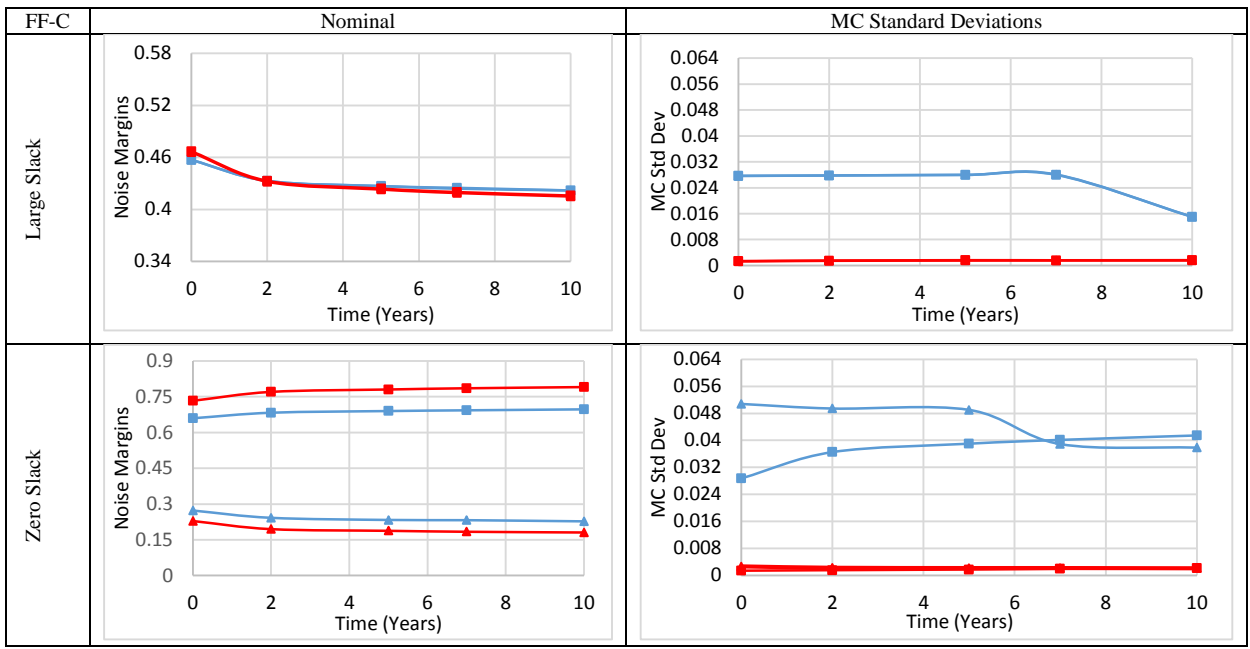

Fig. 11 $V_{writeH}$, $V_{writeL}$ nominal values and standard deviations for flip-flop-C

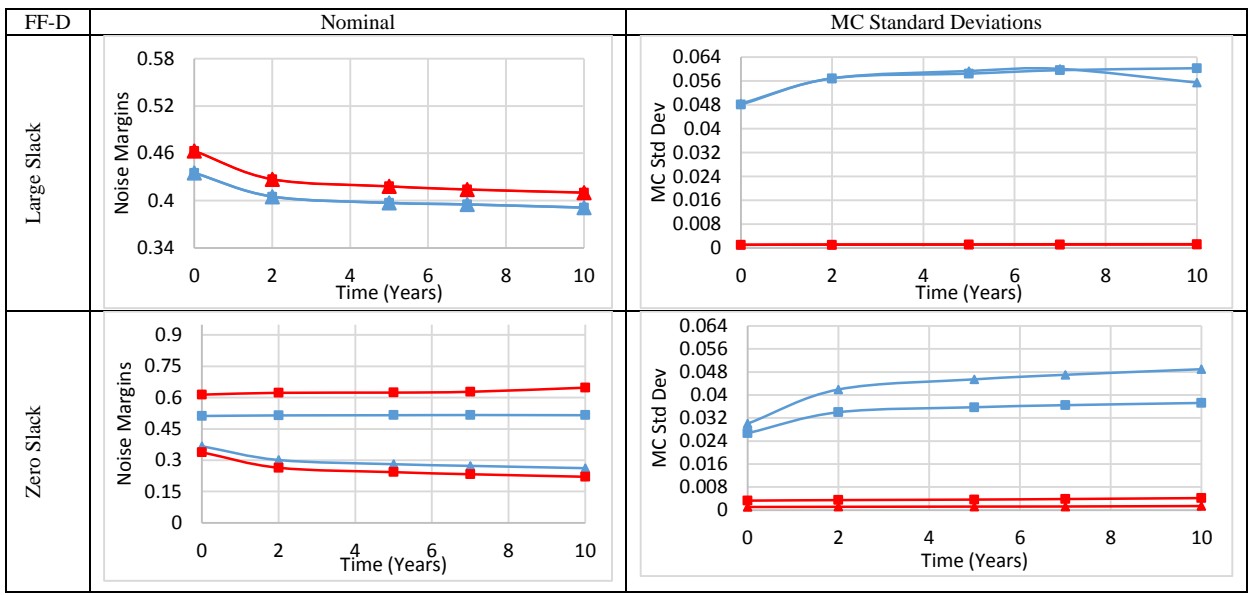

Fig. 12 $V_{writeH}$, $V_{writeL}$ nominal values and standard deviations for flip-flop-D

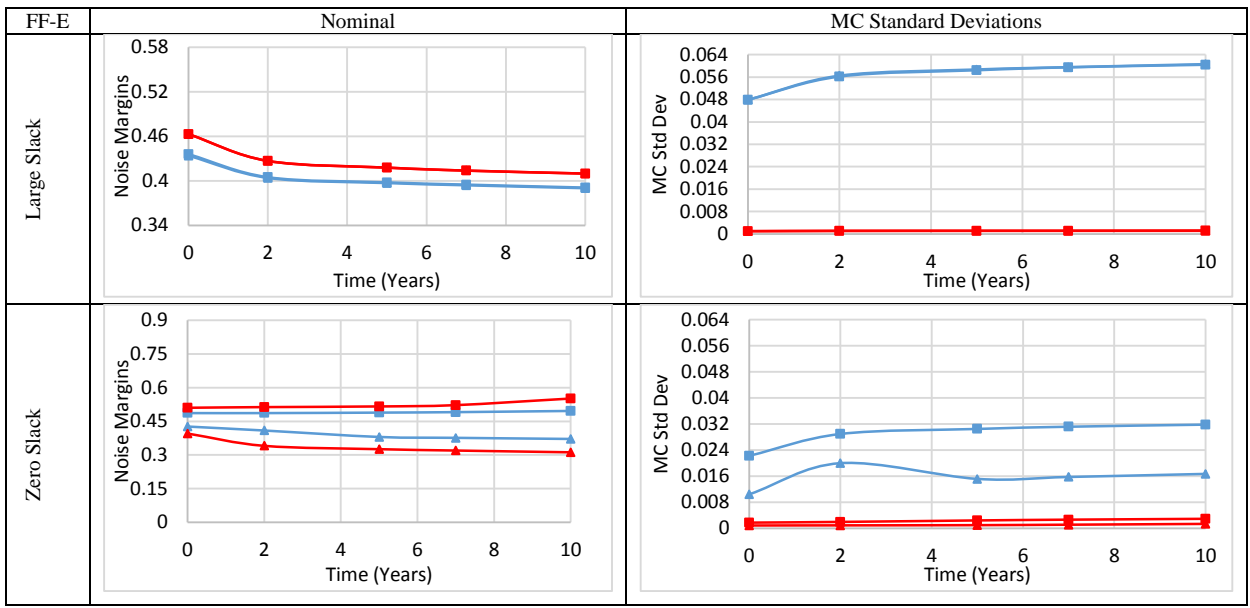

Fig. 13 $V_{writeH}$, $V_{writeL}$ nominal values and standard deviations for flip-flop-E

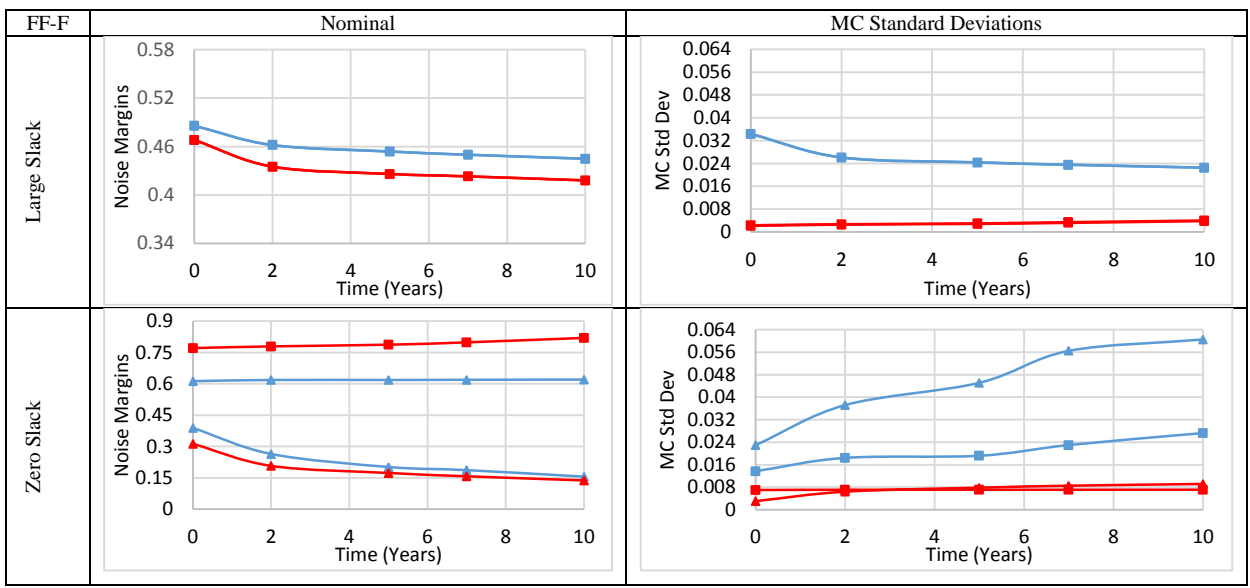

Fig. 14 $V_{writeH}$, $V_{writeL}$ nominal values and standard deviations for flip-flop-F

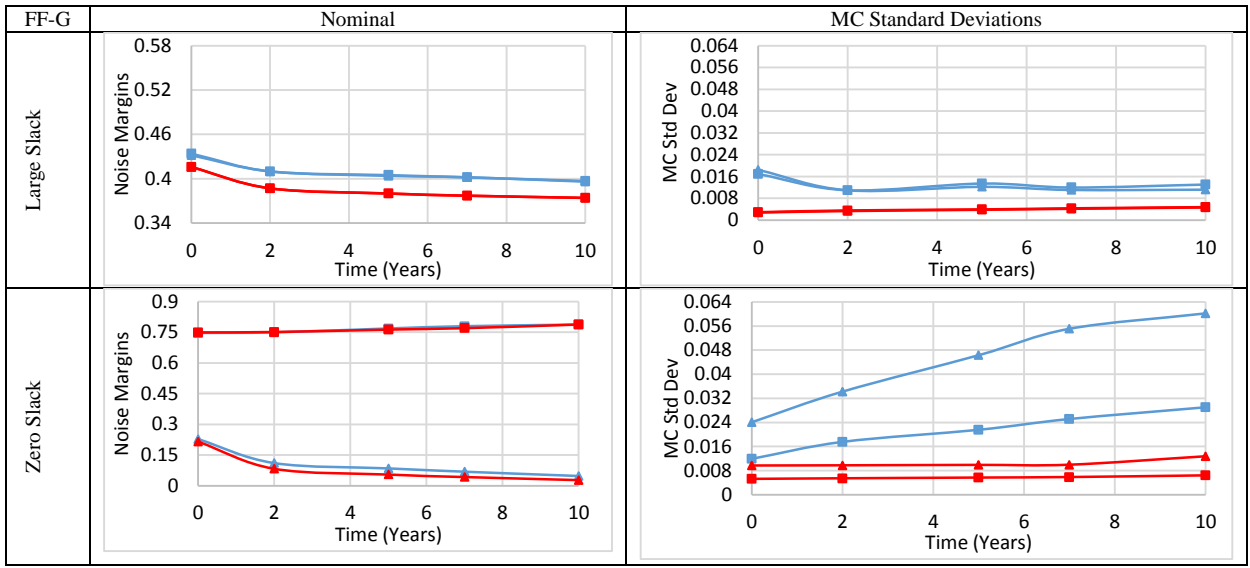

Fig. 15 $V_{writeH}$, $V_{writeL}$ nominal values and standard deviations for flip-flop-G

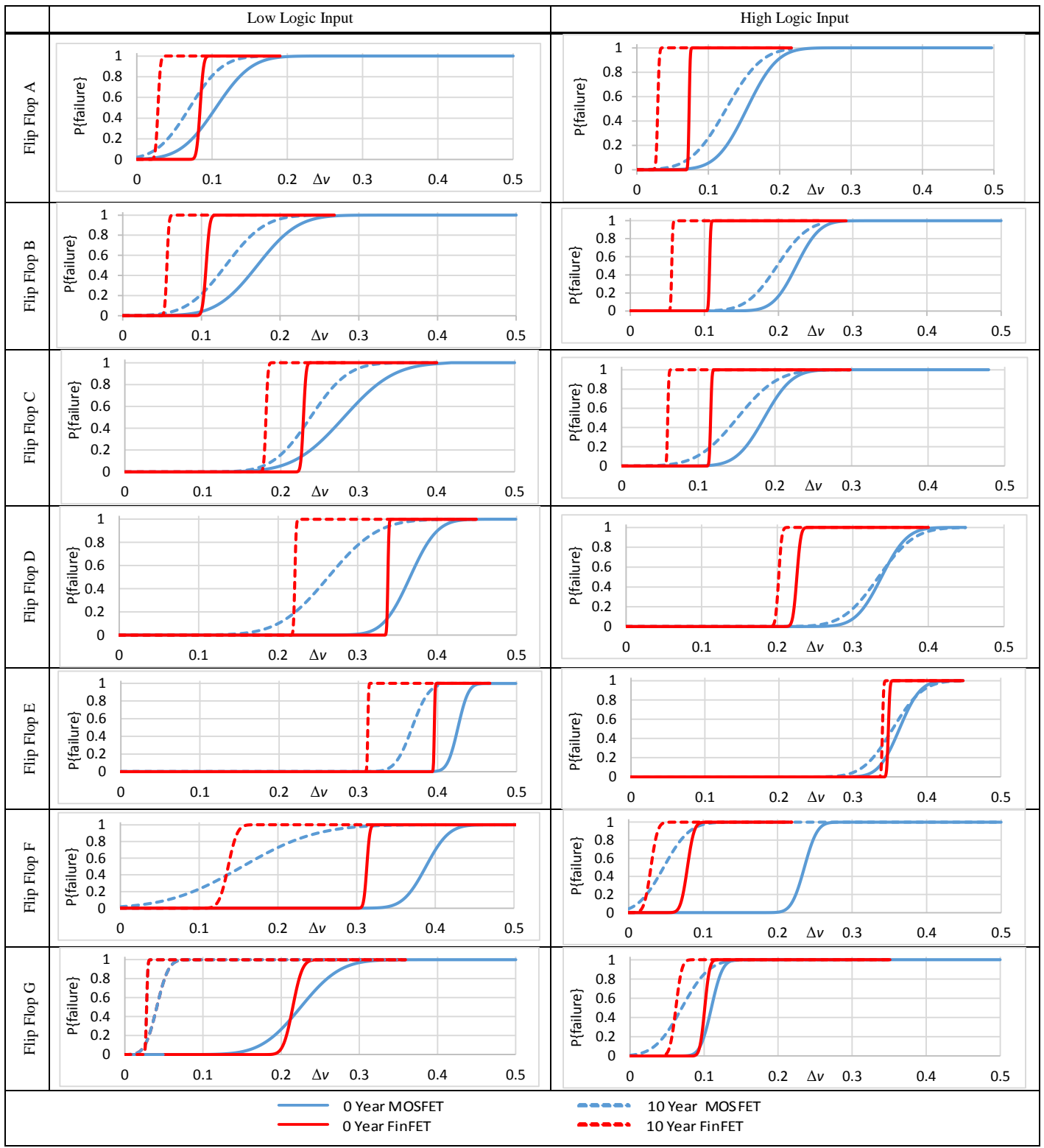

Fig. 16. Write Failure Probability of the Flip-Flop cells vs input voltage shift $\Delta v$ (on the x axis).